\documentstyle[preprint,aps]{revtex}
\begin{document}
\tightenlines
%\draft
%%
%%
\title{Generalization ability of a perceptron with 
non-monotonic transfer function}
%%
%%
%\date{today}
%%
%%
\author{Jun-ichi Inoue$^{\dag}$$^{\P}$, Hidetoshi Nishimori$^{\dag}$ and 
Yoshiyuki Kabashima$^{\ddag}$}
\address{$^{\dag}$Department of Physics, 
Tokyo Institute of Technology, \\
Oh-okayama 
Meguro-ku
Tokyo 152, 
Japan 
}
\address{$^{\ddag}$Department of Computational Intelligence and Systems Science, \\
Interdisciplinary Graduate School of Science and Engineering, 
Tokyo Institute of Technology, Yokohama 226, Japan}
\address{$^{\P}$Laboratory  for  Information Synthesis, 
Brain-Style Information systems, 
RIKEN, \\
Hirosawa 2-1, Wako-shi, Saitama 351-01, Japan
}
\maketitle
\begin{abstract}
We investigate the generalization ability of 
a perceptron with non-monotonic transfer 
function of a reversed-wedge type in on-line mode. 
This network is identical to a parity machine, a multilayer 
network. 
We consider several learning 
algorithms. 
By the perceptron algorithm
the generalization error is shown to
decrease by the ${\alpha}^{-1/3}$-law 
similarly to the case of 
a simple perceptron in a restricted range of
the parameter $a$ characterizing
the non-monotonic transfer function. 
For other values of $a$, the perceptron algorithm 
leads to the state where the weight vector of the
student is just opposite to that of the teacher. 
The Hebbian learning algorithm has a 
similar property; it works only in a limited 
range of the parameter. 
The conventional AdaTron algorithm
does not give a vanishing generalization error 
for any values of $a$. 
We thus introduce a modified 
AdaTron algorithm which yields a 
good performance for all values of $a$. 
We also investigate  the effects of
optimization of the learning 
rate as well as of the learning algorithm. 
Both methods give excellent learning 
curves proportional to ${\alpha}^{-1}$. 
The latter optimization is related to the Bayes statistics and
is shown to yield useful hints to extract maximum amount
of information necessary to accelerate learning processes. 
\end{abstract}
%%%%%%%%%%%%%%%%%%%%%%%%%%%%%%%%%%%%%%%%%%%%%%%%%%%%%%%%%%%%%%%%%%%
%%%%%%%%%%%%%%%%%%%%%%%%%%%%%%%%%%%%%%%%%%%%%%%%%%%%%%%%%%%%%%%%%%%
\section{Introduction}
%%%%%%%%%%%%%%%%%%%%%%%%%%%%%%%%%%%%%%%%%%%%%%%%%%%%%%%%%%%%%%%%%%%
%%%%%%%%%%%%%%%%%%%%%%%%%%%%%%%%%%%%%%%%%%%%%%%%%%%%%%%%%%%%%%%%%%%
%%
%%
In artificial neural networks, the issue of learning from examples 
has been one of the 
most attractive problems \cite{Amari67,Hertz91,Watkin93,Opper95}.
Traditionally emphasis has been put on the 
off-line (or batch) learning.
In the off-line learning scenario, the student 
sees a set of examples (called a training set)
repeatly until the equilibrium is reached. 
This learning scenario 
can be analyzed in the framework of 
equilibrium statistical 
mechanics based on the energy cost function 
which means student's total error for 
a training set or 
on other types of cost 
functions \cite{Griniasti91,Meir92,Kinouchi96}.
However, recently,  several important  
features of learning from 
examples were derived from the paradigm 
of on-line learning.
In the on-line learning scenario, the student 
sees each example only once and 
throws it out, and he never sees it again. 
In other words, at each learning 
stage, the student receives a randomly drawn 
example and is not able to memorize it. 
The most recent example is used for 
modifying the student 
weight vector only by a small amount. 
The on-line learning 
has an advantage over the off-line counterpart  
that it explicitly carries information 
about the current stage of achievement of the student
as a function of the 
training time (which is 
proportional to the 
number of examples). 

During these several years, many interesting 
results have been reported  in relation to the 
on-line learning. 
Among them, the generalization ability 
of multilayer networks is one of the
central problems \cite{Kaba94,Sompo95,Saad95}.
Multilayer neural networks 
are much more powerful machines for 
information representation than 
the simple perceptron. 

Recently, the properties of neural networks 
with a non-monotonic transfer function 
have also been investigated by several authors 
\cite{Morita90,Nishi,Ino,Morita96,Boff93,Monasson94}. 
A perceptron with a non-monotonic transfer function 
has the same input-output 
relations as a multilayer 
neural network called the parity machine. 
This parity machine has one hidden layer
composed of three hidden units 
(the $K=3$ parity machine). 
The output of each unit is represented as 
${\rm sgn}(-u)$, ${\rm sgn}(-a-u)$ and 
${\rm sgn}(a-u)$, where $u\,{\equiv}\,
\sqrt{N}({\bf J}{\cdot}{\bf x})/|{\bf J}|$. 
Here ${\bf J}$ is the $N$-dimensional 
synaptic connection vector and ${\bf x}$ 
denotes the input signal. 
Then the final output of this 
machine is given as the product 
${\rm sgn}(-u){\cdot}{\rm sgn}(-a-u)
{\cdot}{\rm sgn}(a-u)$. 
We regard this final output of the $K=3$ parity 
machine as the output of a perceptron with 
non-monotonic transfer function.
Recently, Engel and Reimers \cite{Engel94} 
investigated the generalization ability of 
this non-monotonic perceptron following 
the off-line learning scenario. 
Their results are summarized as follows; 
For $0<a<\infty$, 
there exists a poor generalization phase 
with a large generalization error. 
As the number of presented patterns increases, a 
good  generalization phase appears after 
a first order phase transition at 
some ${\alpha}$. 
No studies have been made about the present 
system following the 
on-line learning scenario. 
In this paper we study the on-line learning process and the 
generalization ability of this non-monotonic 
perceptron by various learning algorithms. 

This paper is organized as follows. 
In the next section we introduce 
our model system and derive the 
dynamical equations with respect to 
two order parameters for a general learning 
algorithm. 
One is the overlap between the teacher 
and student weight vectors and 
the other is the length 
of the student weight vector. 
In Sec. III, we investigate 
the dynamics of 
on-line learning in the non-monotonic perceptron for 
the conventional perceptron learning and 
Hebbian leaning algorithms. 
We also investigate the asymptotic form of 
the differential equations in both 
small and large ${\alpha}$ 
limits and get 
the asymptotic behavior 
of the generalization error.
In Sec. IV  
we investigate the AdaTron learning algorithm and modify the 
conventional AdaTron algorithm.
In this modification procedure, 
we improve the weight function of 
the AdaTron learning so as to adopt it according to 
the range of $a$.
In Sec. V, we optimize 
the learning rate 
and the general weight 
function appearing in 
the on-line dynamics. 
As the weight function contains 
the variables unknown for 
the student, we average over 
these variables over distribution function 
unknown using the Bayes formula.
Section  VI contains concluding remarks. 
%%
%%
%%%%%%%%%%%%%%%%%%%%%%%%%%%%%%%%%%%%%%%%%%%%%%%%%%%%%
%%%%%%%%%%%%%%%%%%%%%%%%%%%%%%%%%%%%%%%%%%%%%%%%%%%%%%
%%%%%%%%%%%%%%%%%%%%%%%%%%%%%%%%%%%%%%%%%%%%%%%%%%%%%%
%%%%%%%%%%%%%%%%%%%%%%%%%%%%%%%%%%%%%%%%%%%%%%%%%%%%%%
\section{The model system and dynamical equations}
%%%%%%%%%%%%%%%%%%%%%%%%%%%%%%%%%%%%%%%%%%%%%
%%%%%%%%%%%%%%%%%%%%%%%%%%%%%%%%%%%%%%%%%%%%%
%%%%%%%%%%%%%%%%%%%%%%%%%%%%%%%%%%%%%%%%%%%%%%%%%%
%%
%%
We investigate the generalization ability of 
the non-monotonic perceptrons for various 
learning algorithms. 
The student and teacher perceptron are 
characterized by their weight vectors, namely  
${\bf J}{\in}\,{\Re}^{N}$ and 
${\bf B}{\in}\,{\Re}^{N}$ with  
$|{\bf B}|=1$, 
respectively.
For a binary input signal 
${\bf x}{\in}\,\{-1,+1\}^{N}$, 
the output is calculated by the non-monotonic 
transfer function as follows: 
\\
\begin{eqnarray}
T_{a}(v)={\rm sign}\left[v(a-v)(a+v)\right] 
\label{teacher}
\end{eqnarray}
for the teacher and 
\\
\begin{eqnarray}
S_{a}(u)={\rm sign}\left[u(a-u)(a+u)\right]
\label{student}
\end{eqnarray}
for the student, where we define the local fields of 
the teacher and student as 
$v\,{\equiv}\,\sqrt{N}({\bf B}
{\cdot}{\bf x})/|{\bf B}|$ 
and 
$u\,{\equiv}\,\sqrt{N}({\bf J}
{\cdot}{\bf x})/|{\bf J}|$, 
respectively. 
The on-line learning 
dynamics is defined by the following general rule 
for the change of the student vector 
under presentation of the $m$th example; 
\\
\begin{equation}
{\bf J}^{m+1}=
{\bf J}^{m}+f(T_{a}(v),u){\bf x}.
\label{general}
\end{equation}
Well-known examples are the 
perceptron learning, 
$f=-S_{a}(u)\,{\Theta}(-T_{a}(v)S_{a}(u))$, 
the Hebbian learning, 
$f=T_{a}(v)$, 
and the AdaTron learning, 
$f=-ul\,{\Theta}(-T_{a}(v)S_{a}(u))$.

We rewrite the update rule, Eq. (\ref{general}),  of 
$\mbox{\boldmath $J$}$ as a set of 
differential equations introducing 
the dynamical order parameter 
describing the 
overlap between the teacher and student 
weight vectors $R^{m}\,{\equiv}\,
({\bf B}{\cdot}
{\bf J}^{m})/
|{\bf J}^{m}|$ and another order 
parameter describing the norm of the 
student weight vector $l^{m}\,{\equiv}\, 
|{\bf J}^{m}|/\sqrt{N}$. 
By taking the overlap of both sides of 
Eq. (\ref{general}) with 
${\bf B}$ and by squaring 
both sides of the same equation, 
we obtain the dynamical equations 
in the limit of large $m$ and $N$ 
keeping ${\alpha}\,{\equiv}\,m/N$ finite as 
\\
\begin{equation}
\frac{d l}{d \alpha}=\frac{1}{2l}
{\ll}f^{2}(T_{a}(v),u)+2f(T_{a}(v),u)\,ul{\gg}
\label{dldagen}
\end{equation}
and
\\
\begin{equation}
\frac{d R}{d \alpha}=\frac{1}{l^{2}}
{\ll}-\frac{R}{2}f^{2}(T_{a}(v),u)-(Ru-v)f(T_{a}(v),u)l{\gg}. 
\label{drdagen}
\end{equation}
Here ${\ll}\cdots{\gg}$ denotes 
the average over the randomness of inputs 
\begin{equation}
{\ll}\cdots{\gg}\,{\equiv}\,\int\int
dudv(\cdots)P_{R}(u,v)
\label{ave1}
\end{equation}
with
\begin{equation}
P_{R}(u,v)\,{\equiv}\,\frac{1}{2\pi\sqrt{1-R^{2}}}
\,{\exp}\left(-\frac{(u^{2}+v^{2}-2Ruv)}{2(1-R^{2})}\right).
\label{dist1}
\end{equation}
As we are interested in the typical behavior 
under our training algorithm, 
we have averaged 
both sides of Eqs.  
(\ref{dldagen}) and (\ref{drdagen})
over all possible 
instances of examples. 
The Gaussian distribution (\ref{dist1}) has been 
derived from the central limit theorem. 

The generalization error, which is the 
probability of disagreement between the teacher and the 
trained student, is represented as 
${\epsilon}_{g}={\ll}{\Theta}(-T_{a}(v)S_{a}(u)){\gg}$. 
After simple calculations, we obtain 
the generalization error as 
\begin{eqnarray}
E(R){\equiv}{\epsilon}_{g} & = & 2\int_{a}^{\infty}
Dv\,H\left(\frac{a+Rv}{\sqrt{1-R^{2}}}\right) 
+2\int_{a}^{\infty}
Dv\,H\left(\frac{-(a-Rv)}{\sqrt{1-R^{2}}}\right) \nonumber \\
\mbox{} & + & 2\int_{0}^{a}
Dv\,H\left(\frac{Rv}{\sqrt{1-R^{2}}}\right)-
2\int_{a}^{\infty}
Dv\,H\left(\frac{Rv}{\sqrt{1-R^{2}}}\right) \nonumber \\
\mbox{} & - & 2\int_{0}^{a}
Dv\,H\left(\frac{a+Rv}{\sqrt{1-R^{2}}}\right) 
+2\int_{0}^{a}
Dv\,H\left(\frac{a-Rv}{\sqrt{1-R^{2}}}\right) 
\label{e2} 
\end{eqnarray}
where we have set $H(x)=\int_{x}^{\infty}Dt$ with 
$Dt\,{\equiv}\,dt\,{\exp}(-t^{2}/2)/\sqrt{2\pi}$. 

We would like to emphasize that the 
generalization error 
obtained above (\ref{e2}) is independent of 
the specific learning algorithm. 
In Fig. 1, we plot $E(R)={\epsilon}_{g}$ 
for several values of $a$. 
This figure tells us that the student can 
acquire a perfect generalization ability 
if he is trained so that $R$ converges to $1$ for 
all values of $a$. 
We have confirmed also analytically that $E(R)$ 
is a monotonically decreasing function of $R$ for any value of $a$.
%%%%%%%%%%%%%%%%%%%%%%%%%%%%%%%%%%%%%%%%%%%%%%%%%%%%%%%%%%%%%%%%%%%%
%%%%%%%%%%%%%%%%%%%%%%%%%%%%%%%%%%%%%%%%%%%%%%%%%%%%%%%
%%%%%%%%%%%%%%%%%%%%%%%%%%%%%%%%%%%%%%%%%%%%%%%%%%%%%%%%%%
\section{Hebbian and Perceptron learning algorithms}
%%%%%%%%%%%%%%%%%%%%%%%%%%%%%%%%%%%%%%%%%%%%%%%%%%%%%%%%%%
%%%%%%%%%%%%%%%%%%%%%%%%%%%%%%%%%%%%%%%%%%%%%%%%%%%%%%%%%%%%
%%%%%%%%%%%%%%%%%%%%%%%%%%%%%%%%%%%%%%%%%%%%%%%%%%%%%%%%%%%%%
\subsection{Hebbian learning}
%%%%%%%%%%%%%%%%%%%%%%%%%%%%%%%%%%%%%%%%%%%%%%%%%%%%%%%%%%%%%
%%%%%%%%%%%%%%%%%%%%%%%%%%%%%%%%%%%%%%%%%%%%%%%%%%%%%%%%%%%%
We first investigate the performance of the 
on-line Hebbian 
learning $f=T_{a}(v)$. 
We get the differential equations for $l$ and $R$ as follows 
\\
\begin{eqnarray}
\frac{d l}{d \alpha} & = & \left[
\frac{1}{2}+\frac{2R}{\sqrt{2\pi}}
(1-2{\Delta})l\right]/l 
\label{dldahebb1} \\
\mbox{}\frac{d R}{d \alpha} & = & \left[
-\frac{R}{2}\frac{2}{\sqrt{2\pi}}
(1-2{\Delta})(1-R^{2})l\right]/l^{2}. 
\label{drdahebb1}
\end{eqnarray}
To determine whether or not $R$ 
increases with ${\alpha}$ according to $a$, 
we approximate the differential 
equation for $R$ around 
$R=0$ as 
\begin{equation}
\frac{d R}{d \alpha}=
\frac{2}{\sqrt{2\pi}}
(1-2{\Delta})\frac{1}{l^{2}}.
\end{equation}
Therefore we use 
$R=1-{\varepsilon}$ for $a>a_{c}\,{\equiv}\,\sqrt{2{\log}2}$ and 
$R={\varepsilon}-1$ for $a<a_{c}$. 
When $a>a_{c}$,  we obtain 
\begin{equation}
{\epsilon}_{g}=\frac{1}{\sqrt{2\pi}}
\frac{1+2{\Delta}}{1-2{\Delta}}
\frac{1}{\sqrt{\alpha}}
\end{equation}
and 
\begin{equation}
l=\sqrt{\frac{2}{\pi}}
(1-2{\Delta}){\alpha}.
\end{equation}
On the other hand, for $a<a_{c}$ we obtain 
\begin{equation}
{\epsilon}_{g}=1+\frac{1}{\sqrt{2\pi}}
\frac{1+2{\Delta}}{1-2{\Delta}}
\frac{1}{\sqrt{\alpha}}
\end{equation}
and 
\begin{equation}
l=-\sqrt{\frac{2}{\pi}}(1-2{\Delta}){\alpha}.
\end{equation}
We see that the Hebbian learning algorithms lead to the state 
$R=-1$ for $a<a_{c}$.
%%
%%
%%

%%%%%%%%%%%%%%%%%%%%%%%%%%%%%%%%%%%%%%%%%%%%%%%%%%%%%%%%%%%%%
\subsection{Perceptron learning}
%%%%%%%%%%%%%%%%%%%%%%%%%%%%%%%%%%%%%%%%%%%%%%%%%%%%%%%%%%%
%%
%%
We next investigate the 
on-line perceptron 
learning $f=-S_{a}(u)\,{\Theta}(-T_{a}(v)S_{a}(u))$ 
by solving the next differential equations numerically; 
\begin{eqnarray}
\frac{d l}{d \alpha} & = & [\frac{1}{2}E(R)-F(R)\,l
]/l
\label{dldaper} \\
\mbox{}\frac{d R}{d \alpha} & = & [
-\frac{1}{2}E(R)R+(F(R)R-G(R))l
]/l^{2}
\label{drdaper}
\end{eqnarray}
where $F(R)={\ll}{\Theta}(-T_{a}(v)S_{a}(u))S_{a}(u)u{\gg}$ and 
$G(R)={\ll}{\Theta}(-T_{a}(v)S_{a}(u))S_{a}(u)v{\gg}$. 
Using the distribution (\ref{dist1}) 
we can rewrite these functions as 
\begin{equation}
F(R)=\frac{(1-R)}{\sqrt{2\pi}}(1-2{\Delta})
\label{frper} 
\end{equation}
and
\begin{equation}
G(R)=-F(R)
\label{grper}
\end{equation}
where ${\Delta}\,{\equiv}\,{\rm exp}(-a^{2}/2)$. 
In Fig. 2 we plot the change of $R$ and $l$ as learning 
proceeds under various initial conditions for the 
case of $a=\infty$.
We see that the 
student can reach the perfect generalization 
state $R=1$ for any initial condition.
The $R$-$l$ flow in the opposite limit $a=0$ 
is shown in Fig. 3.
Apparently, for this case 
the student 
reaches the state with the weight 
vector opposite to the teacher, 
$R=-1$, after 
an infinite number of 
patterns are presented. 
%%
%%
%%
%%
%%
%%
%%%%%%%%%%%%%%%%%%%%%%%%%%%%%%%%%%%%%%%%%%%%%%%%%%%%%%%%%%%%%%
From Eqs. (\ref{teacher}) and (\ref{student}), 
we should notice that the case of $a=0$ is essentially 
different from the case of a simple perceptron. 
%%%%%%%%%%%%%%%%%%%%%%%%%%%%%%%%%%%%%%%%%%%%%%%%%%%%%%%%%%%%%%%

Since the two limiting cases, 
$a=\infty$ and $a=0$, follow different types 
of behavior, it is necessary to check what happens 
in the intermediate region. 
For this purpose, we first investigate 
the asymptotic behavior of 
the solution of Eqs.  (\ref{dldaper}) and (\ref{drdaper}) 
near $R={\pm}1$ for large $\alpha$. 
Using the notation $R=1-{\varepsilon}$, 
${\varepsilon}{\rightarrow}0$, the 
asymptotic forms of $E(R)$, $F(R)$ and $G(R)$ are 
found to be 
\begin{eqnarray}
E(R) & {\simeq} & \frac{\sqrt{2{\varepsilon}}}{\pi}
(1+2{\Delta})
\label{asme1} \\
\mbox{}F(R) & {\simeq} & \frac{\varepsilon}{\sqrt{2\pi}}(1-2{\Delta})
\label{asmf1} \\
\mbox{}G(R) & {\simeq} & -\frac{\varepsilon}{\sqrt{2\pi}}(1-2{\Delta})
\label{asmg1}.
\end{eqnarray}
Substituting these expressions into the differential 
equations (\ref{dldaper}) and (\ref{drdaper}),  we obtain 
\begin{eqnarray}
{\varepsilon} & = & \left[
\frac{(1+2{\Delta})}{3\sqrt{2}(1-2{\Delta})^{2}}
\right]^{2/3}
\,{\alpha}^{-2/3}
\label{asmeps1} \\
\mbox{}l & = & 
\frac{1}{2\sqrt{\pi}}\left(
\frac{1+2{\Delta}}{1-2{\Delta}}
\right)
\left[
\frac{3\sqrt{2}(1-2{\Delta})^{2}}{(1+2{\Delta})}
\right]^{1/3}
\,{\alpha}^{1/3}.
\label{asml1}
\end{eqnarray}
Therefore, the generalization error is obtained from  
Eq. (\ref{asme1}) as 
\begin{equation}
{\epsilon}_{g}=(1+2{\Delta})\frac{\sqrt{2}}{\pi}
\left[
\frac{(1+2{\Delta})}{3\sqrt{2}(1-2{\Delta})^{2}}
\right]^{1/3}\,{\alpha}^{-1/3}. 
\label{asmge1}
\end{equation}

The asymptotic form of $l$, Eq. (\ref{asml1}), 
shows that ${\Delta}$ should satisfy 
$2{\Delta}<1$ or $a > a_{c}$. 
The assumption of $R=1-{\varepsilon}$ with 
${\varepsilon}{\rightarrow}0$ thus fails if 
$a < a_{c}$. 
This fact can be verified from Eq.  (\ref{drdaper}) 
expanded around $R=0$ as 
\begin{equation}
\frac{d R}{d \alpha}\,{\simeq} \frac{2}{\sqrt{2\pi}}
(1-2{\Delta})\frac{1}{l^{2}}.
\label{simzero}
\end{equation}
For $a<a_{c}$, 
$R$ decreases with ${\alpha}$.
Therefore, we use the relation 
$R={\varepsilon}-1, {\varepsilon}{\rightarrow}0$,  instead of 
$R=1-{\varepsilon}$ for $a<a_{c}$.
We then find the asymptotic form of 
the generalization error as 
\begin{equation}
{\epsilon}_{g}=1+\left[
\frac{1+2{\Delta}}{1-2{\Delta}}
\right]
\frac{1}{\sqrt{2\pi{\alpha}}}
\label{asmge2}
\end{equation}
and $l$ goes to infinity as 
\begin{equation}
l=-\frac{2}{\sqrt{2\pi}}
(1-2{\Delta}){\alpha}.
\label{asml2}
\end{equation}
These two results,  Eqs.  
(\ref{asmge1}) and (\ref{asmge2}), 
confirm the difference in the asymptotic behaviors 
between the two cases of $a=0$ and $a=\infty$. 

We have found that the Hebbian and the conventional 
perceptron learning algorithms 
lead to the state $R=-1$ for $a<a_{c}=\sqrt{2{\log}2}$.
%%%%%%%%%%%%%%%%%%%%%%%%%%%%%%%%%%%%%%%%%%%%%%%%%
%    This part should be added to section III   %
%%%%%%%%%%%%%%%%%%%%%%%%%%%%%%%%%%%%%%%%%%%%%%%%%   
%%
%%
This anti-learning effect may be understood 
as follows. 
If the student perceptron has learned only one example by 
the Hebb rule, 
\begin{equation}
{\bf J}=T_{a=0}(v){\bf x}.
\label{tenehebb}
\end{equation}
Then the output of the student for the same example is 
\begin{eqnarray}
S_{a=0}(u) & = & -{\rm sgn}(u) \nonumber \\
\mbox{} & = & -{\rm sgn}\left({\bf J}{\cdot}{\bf x}\right) \nonumber \\
\mbox{} & = & -T_{a=0}(v). 
\label{sonehebb}
\end{eqnarray}
This relation indicates the anti-learning effect for the 
$a=0$ case. Similar 
analysis holds for the perceptron learning. 
%%%%%%%%%%%%%%%%%%%%%%%%%%%%%%%%%%%%%%%%%%%%%%%%%%%%%%%%%
%%%%%%%%%%%%%%%%%%%%%%%%%%%%%%%%%%%%%%%%%%%%%%%%%%%%%%%%%%%
%%%%%%%%%%%%%%%%%%%%%%%%%%%%%%%%%%%%%%%%%%%%%%%%%%%%%%%%%%%%%
\subsection{Generalized perceptron learning}
%%%%%%%%%%%%%%%%%%%%%%%%%%%%%%%%%%%%%%%%%%%%%%%%%%%%%%%%%%%%
%%%%%%%%%%%%%%%%%%%%%%%%%%%%%%%%%%%%%%%%%%%%%%%%%%%%%%%%%%%%
%%
%%
%%
%%
In this section, 
we introduce a multiplicative factor 
$|u|^{\gamma}$ in front of 
the perceptron learning function, 
$f=-|u|^{\gamma}{\Theta}
(-T_{a}(v)S_{a}(u))S_{a}(u)$, and 
investigate how the 
generalization ability depends on the 
parameter $\gamma$. 
In particular, we are interested in whether or not 
an optimal value of $\gamma$ exists. 
The learning dynamics is therefore 
\begin{equation}
{\bf J}^{m+1}={\bf J}^{m}
-|u|^{\gamma}S_{a}(u)
{\Theta}(-T_{a}(v)S_{a}(u))
{\bf x}. 
\label{genpercep}
\end{equation}
The case of 
$\gamma=0$ corresponds to the conventional 
perceptron learning algorithm. 
%%
%%
%%%%%%%%%%%%%%%%%%%%%%%%%%%%%%%%%%%%%%%%%%%%%%%%%%%%%%%%%%%%%%
On the other hand, the case of $\gamma=1$ and 
$a{\rightarrow}\infty$ corresponds to 
the conventional AdaTron learning.
%%%%%%%%%%%%%%%%%%%%%%%%%%%%%%%%%%%%%%%%%%%%%%%%%%%%%%%%%%%%%
%%
%%
Using the above learning dynamics, 
we obtain the differential equations with 
respect to $l$ and $R$ as  
\begin{eqnarray}
\frac{d l}{d \alpha} & = & \frac{1}{l}
\left[
\frac{E_{G}(R)}{2}-lF_{G}(R)
\right]
\label{dlda3} \\
\mbox{}\frac{d R}{d \alpha} & = & \frac{1}{l^{2}}
\left[
-\frac{R}{2}E_{G}(R)+(
F_{G}(R)R-G_{G}(R))l
\right], 
\label{drda3}
\end{eqnarray}
where $E_{G}(R)$, $F_{G}(R)$ and $G_{G}(R)$ are 
represented as 
\begin{eqnarray}
E_{G}(R) & {\equiv} & {\ll}u^{2\gamma}{\Theta}(-T_{a}(v)S_{a}(u)){\gg},
\label{eg} \\
\mbox{}F_{G}(R) & {\equiv} & {\ll}
|u|^{{\gamma}+1}{\Theta}(-T_{a}(v)S_{a}(u))S_{a}(u){\gg}
\label{fg} 
\end{eqnarray}
and 
\begin{equation}
\mbox{}G_{G}(R){\equiv}{\ll}
|u|^{\gamma}{\Theta}(-T_{a}(v)S_{a}(u))S_{a}(u)v{\gg}.
\label{gg}
\end{equation}

Let us first investigate the 
behavior of the $R$-$l$ flow near $R=0$.  
When $R$ is very small, the right-hand side of 
Eq. (\ref{drda3}) is found to be a 
$\gamma$-dependent constant: 
\begin{equation}
\frac{d R}{d \alpha}=\frac{2^{\frac{1}{2}(\gamma-1)}}
{{\pi}l}{\Gamma}
\left(\frac{\gamma}{2}+\frac{1}{2}
\right)(1-2{\Delta}),
\label{asm}
\end{equation}
where ${\Gamma}(x)$ is the gamma 
function.  
As the right hand side of Eq.  (\ref{asm}) is positive 
for any $\gamma$  
as long as $a$ satisfies $a>a_{c}$, 
$R$ increases around $R=0$ only for this range of $a$. 
Thus the generalized perceptron learning algorithm 
succeeds in reaching the desired state $R=1$, 
not the opposite one $R=-1$, 
only for $a>a_{c}$, similarly to the conventional 
perceptron learning. 
Therefore, in this section we restrict our 
analysis to the case of $a>a_{c}$ 
and investigate how  the learning curve 
changes according to the value of $\gamma$.

Using the notation $R=1-{\varepsilon}$ 
(${\varepsilon}{\rightarrow}0$), we obtain the 
asymptotic forms of $E_{G}$, $F_{G}$ and $G_{G}$ as 
follows. 
\begin{eqnarray}
E_{G} & {\simeq} & c_{1}{\varepsilon}^{\gamma+\frac{1}{2}}
+c_{2}{\varepsilon}^{\frac{1}{2}}
\label{asmeg1} \\
F_{G} & {\simeq} & c_{3}{\varepsilon}^{1+\frac{\gamma}{2}}
-c_{4}{\varepsilon}
\label{asmfg1} \\
G_{G} & {\simeq} & -\frac{c_{3}}{\gamma+1}{\varepsilon}^{1+\frac{\gamma}{2}}
+c_{4}{\varepsilon}
\label{asmgg1}
\end{eqnarray}
where $c_{1}\,{\equiv}\,{2^{2\gamma+1/2}{\Gamma}(\gamma+1)}/
{\pi(2\gamma+1)}$, 
$c_{2}\,{\equiv}\,{4a^{2\gamma}{\Delta}}/{\sqrt{2}\pi}$, 
$c_{3}\,{\equiv}\,{2^{\gamma+3/2}{\Gamma}(\frac{\gamma}{2}
+\frac{3}{2})}/{\pi(\gamma+2)}$ and 
$c_{4}\,{\equiv}\,{2{\Delta}a^{\gamma}}/{\sqrt{2\pi}}$.
We first investigate the case of 
${\Delta}{\neq}0$ (finite $a$), namely, $c_{2},c_{4}{\neq}0$. 
The differential 
equations (\ref{dlda3}) and (\ref{drda3}) 
are rewritten in terms of ${\varepsilon}$ and 
${\delta}=1/l$ as
\begin{eqnarray}
\frac{d \delta}{d \alpha} & = & -\frac{{\delta}^{3}}
{2}
\left[
c_{1}{\varepsilon}^{\gamma+\frac{1}{2}}+c_{2}{\varepsilon}^{\frac{1}{2}}
\right]
+
{\delta}^{2}
\left[
c_{3}{\varepsilon}^{1+\frac{\gamma}{2}}
-c_{4}{\varepsilon}
\right]
\label{asmdlda2} \\
\frac{d \varepsilon}{d \alpha} & = & 
\frac{{\delta}^{2}}{2}
\left[
c_{1}{\varepsilon}^{\gamma+\frac{1}{2}}
+c_{2}{\varepsilon}^{\frac{1}{2}}
\right]
-{\delta}
\left[
\left(
\frac{2+\gamma}{1+\gamma}
\right)
c_{3}{\varepsilon}^{1+\frac{\gamma}{2}}
-2c_{4}{\varepsilon}
\right].
\label{asmdrda2}
\end{eqnarray}
As $\gamma=0$ corresponds to the perceptron 
learning, we now assume $\gamma{\neq}0$. 
When $\gamma > 0$, the terms  containing 
$c_{1}$ and $c_{3}$ can be neglected in the 
leading order. 
Dividing Eq. (\ref{asmdlda2}) by Eq. (\ref{asmdrda2}), 
we obtain 
\begin{equation}
\frac{d \delta}{d \varepsilon}
=\frac{{\delta}
\left[
-c_{2}{\delta}{\varepsilon}^{1/2}/2
-c_{4}{\varepsilon}
\right]
}
{
\left[
c_{2}{\delta}{\varepsilon}^{1/2}/2
+2c_{4}{\varepsilon}
\right]
}.
\label{ddde}
\end{equation}
If we assume ${\delta}{\varepsilon}^{1/2}{\gg}{\varepsilon}$ 
or ${\delta}{\varepsilon}^{1/2}{\ll}{\varepsilon}$, 
Eq. (\ref{ddde}) is solved as ${\delta}={\exp}(-\varepsilon)$, 
which is in contradiction to the assumption 
$|{\delta}|{\ll}1$. 
Thus, we set 
\begin{equation}
{\delta}=-\frac{4c_{4}}{c_{2}}{\varepsilon}^{1/2}
+b{\varepsilon}^{c}
\label{sol1}
\end{equation}
and determine $b$ and $c(>1/2)$. 
Substituting (\ref{sol1}) into (\ref{ddde}), we find 
$b=8c_{4}/c_{2}$ ($c_{2},c_{4}> 0$)and $c=3/2$. 
The negative value of 
${\delta}=1/l$ is not acceptable 
and we conclude that $R$ does not 
approach $1$ when $\gamma > 0$.

Next we investigate the case of 
${\gamma}<0$. 
Using the same technique as in the case of 
$\gamma>0$, we obtain 
\begin{eqnarray}
{\varepsilon} & = & 
\left[
\frac{c_{1}(1+\gamma)(1-{\gamma}^{2})}
{6c_{3}^{2}(\gamma+2)}
\right]^{\frac{2}{3}}
{\alpha}^{-\frac{2}{3}}, \\
{\delta} & = & \frac{2c_{3}}{c_{1}}
\left(
\frac{\gamma+2}{\gamma+1}
\right)
{\varepsilon}^{\frac{1}{2}(1-\gamma)}
-\frac{4c_{3}}{c_{1}(1-{\gamma}^{2})}
{\varepsilon}^{\frac{1}{2}(3-\gamma)}
\end{eqnarray}
and 
\begin{eqnarray}
{\epsilon}_{g} & = & \frac{\sqrt{2}}{\pi}(1+2{\Delta})
\left[
\frac{c_{1}(1+\gamma)(1-{\gamma}^{2})}
{6c_{3}^{2}(\gamma+2)}
\right]^{\frac{1}{3}}
{\alpha}^{-\frac{1}{3}} \nonumber \\
\mbox{} & {\equiv} & \frac{\sqrt{2}}{\pi}(1+2{\Delta})
f(\gamma){\alpha}^{-\frac{1}{3}}.
\end{eqnarray}
We notice that 
$\gamma$ should satisfy $-1<\gamma<0$, 
because the prefactor of the leading term of 
$\delta$, namely, 
$(2c_{3}/c_{1})(\gamma+2)/(\gamma+1)$, 
must be positive. 
As the prefactor of the generalization error 
increases monotonically from $\gamma=-1$ to 
$\gamma=0$, 
we obtain a smaller generalization 
error for $\gamma$ closer to $-1$. 

Next we investigate the case of 
$a{\rightarrow}\infty$, namely 
$c_{2},c_{4}=0$. 
We first assume  
$l{\rightarrow}l_{0}$ 
in the limit of ${\alpha}{\rightarrow}\infty$. 
In this solution, ${d l}/{d \alpha}=0$ 
should be satisfied asymptotically. 
Then, from Eq. (\ref{asmdlda2}), the two terms 
${\varepsilon}^{{\gamma}+\frac{1}{2}}$ and 
${\varepsilon}^{1+\frac{\gamma}{2}}$ should be equal to 
each other, namely, ${\varepsilon}^{{\gamma}+\frac{1}{2}}
={\varepsilon}^{1+\frac{\gamma}{2}}$, 
which leads to $\gamma=1$.
The learning dynamics (\ref{genpercep}) with 
$a{\rightarrow}\infty$  and $\gamma=1$ is nothing but 
the AdaTron learning which has already been investigated 
in detail \cite{Ino2}.
The result for the generalization error is
\begin{equation}
{\epsilon}_{g}=\frac{3}{2\alpha}, 
\end{equation}
if we choose $l_{0}$ as $l_{0}=1/2$, and 
\begin{equation}
{\epsilon}_{g}=\frac{4}{3\alpha}. 
\end{equation}
if we optimize $l_{0}$ to minimize the 
generalization error.

We next assume $l{\rightarrow}\infty$ 
as ${\alpha}{\rightarrow}\infty$. 
It is straightforward to see that 
${\varepsilon}$ has the 
same asymptotic form as in the case 
of ${\Delta}{\neq}0$ and ${\gamma}<0$. 
Thus we have 
\begin{equation}
{\epsilon}_{g}=\frac{\sqrt{2}}{\pi}f_{2}(\gamma){\alpha}^{-\frac{1}{3}},
\end{equation}
where $f_{2}(\gamma)$ is defined as 
\begin{equation}
f_{2}(\gamma)=\left[
\frac{{\pi}(1+\gamma)(1-{\gamma}^{2}){\Gamma}(\gamma+1)}
{6{\cdot}2^{5/2}{\Gamma}^{2}(\frac{\gamma}{2}+\frac{1}{2})}
\right]^{\frac{1}{3}}
\label{f2}
\end{equation}
and $\gamma$ can take any value within 
$-1<\gamma<0$. 

From the above 
analysis, we conclude that 
the student can get the generalization ability 
${\alpha}^{-1}$ if and only if $a{\rightarrow}\infty$  and 
${\gamma}=1$ (AdaTron).
For other cases  
the generalization error behaves as ${\alpha}^{-1/3}$, 
the same functional form as in the case of the 
conventional perceptron learning, 
as long as the student can obtain a vanishing residual 
error. 
Therefore the learning 
curve has universality in the sense that 
it does not depend on the detailed value of the 
parameter $\gamma$.
%%
%%
%%
%%
%%%%%%%%%%%%%%%%%%%%%%%%%%%%%%%%%%%%%%%%%%%%%%%%%%%%%%%%%%%
%%%%%%%%%%%%%%%%%%%%%%%%%%%%%%%%%%%%%%%%%%%%%%%%%%%%%%%%%%%%%
\section{AdaTron learning algorithm}
%%%%%%%%%%%%%%%%%%%%%%%%%%%%%%%%%%%%%%%%%%%%%%%%%%%%%%%%%%%%%%
\subsection{AdaTron learning}
%%%%%%%%%%%%%%%%%%%%%%%%%%%%%%%%%%%%%%%%%%%%%%%%%%%%%%%%%%%%%
%%
%%
In this subsection, we investigate 
the generalization performance of the conventional 
AdaTron learning $f=-ul{\Theta}(-T_{a}(v)S_{a}(u))$ \cite{Biehl94}. 
The differential equations for $l$ and $R$ 
are given as follows: 
\begin{eqnarray}
\frac{d l}{d \alpha} & = & -\frac{l}{2}E_{\rm Ad}(R)
\label{dldaadat} \\
\frac{d R}{d \alpha} & = & \frac{R}{2}E_{\rm Ad}(R)-G_{\rm Ad}(R)
\label{drdaadat}
\end{eqnarray}
where $E_{\rm Ad}(R)={\ll}u^{2}{\Theta}(-T_{a}(v)S_{a}(u)){\gg}$ 
and $G_{\rm Ad}(R)={\ll}uv{\Theta}(-T_{a}(v)S_{a}(u)){\gg}$. 
After simple calculations, we obtain 
\begin{eqnarray}
E_{\rm Ad}(R) & = & 2\left(
\int_{a}^{\infty}+\int_{-a}^{0}
\right)
Du\,u^{2}\,H\left(
\frac{a+Ru}{\sqrt{1-R^{2}}}
\right) \nonumber \\
\mbox{} & + & 
2\left(\int_{0}^{a}+\int_{-\infty}^{-a}
\right)
Du\,u^{2}\,\left[H
\left(
\frac{Ru}{\sqrt{1-R^{2}}}
\right)
-H\left(
\frac{a+Ru}{\sqrt{1-R^{2}}}
\right)
\right]
\label{eadat}
\end{eqnarray}
and
{\small 
\begin{eqnarray}
G_{\rm Ad}(R) & = & E_{\rm Ad}(R)R \hspace{4.0in}\nonumber \\
\mbox{} & + & \frac{4Ra\Delta}{\sqrt{2\pi}}
(1-R^{2})
\left[
H\left(
\frac{a(1+R)}{\sqrt{1-R^{2}}}
\right)
-H\left(
\frac{aR}{\sqrt{1-R^{2}}}
\right)
-H\left(
\frac{a(1-R)}{\sqrt{1-R^{2}}}
\right)
+\frac{1}{2}
\right] \nonumber \\
\mbox{} & + & 
\frac{2(1-R^{2})^{\frac{3}{2}}}{\pi} \nonumber \\
\mbox{}& {\times} & {\Biggr [}
{\Delta}{\exp}\left[-\frac{a^{2}R^{2}}{2(1-R^{2})}\right]
-{\Delta}{\exp}\left[-\frac{a^{2}(1+R)^{2}}{2(1-R^{2})}\right]
-{\Delta}{\exp}\left[-\frac{a^{2}(1-R)^{2}}{2(1-R^{2})}\right] \nonumber \\
\mbox{} & + & {\exp}\left[-\frac{a^{2}}{2(1-R^{2})}\right]-\frac{1}{2} 
{\Biggr ]}
\label{gadat}
\end{eqnarray}
}
At first, we check the behavior of $R$ around $R=0$. 
Evaluating the differential equation (\ref{drdaadat}) 
around $R=0$, we obtain 
\begin{equation}
\frac{d R}{d \alpha}=\frac{4}{\pi}
\left({\Delta}-\frac{1}{2}\right)^{2}
\label{drdaadat0}
\end{equation}
From this result we find that for any value of $a$, 
the flow of $R$ increases around $R=0$. 
In Fig. 4, we display the flows 
in the $R$-$l$ plane for several values of $a$ 
by numerical integration of Eq. (\ref{drdaadat}). 
This figure indicates that the overlap $R$ increases 
monotonically, but $R$ does not reach the state $R=1$
if $a$ is finite. 
This means that 
the differential equation (\ref{drdaadat}) 
with respect to $R$ has a non-trivial fixed point 
$R=R_{0}$($<1$) if $a<\infty$, which is the solution of 
the non-linear equation $RE_{\rm Ad}(R)=2G_{\rm Ad}(R)$.
Therefore, we conclude that for $a=\infty$ and $a=0$, 
we obtain the generalization error as 
${\epsilon}_{g}\,{\sim}\,{\alpha}^{-1}$, but the 
generalization error converges to 
a finite value exponentially for finite $a$. 
In Fig. 5, we plot  the corresponding generalization error.
%%
%%
%%
%%%%%%%%%%%%%%%%%%%%%%%%%%%%%%%%%%%%%%%%%%%%%%%%%%%%%%%%%%%%%%%%%%%%
%%%%%%%%%%%%%%%%%%%%%%%%%%%%%%%%%%%%%%%%%%%%%%%%%%%%%%%%%%%%%%%%%%%%
%%%%%%%%%%%%%%%%%%%%%%%%%%%%%%%%%%%%%%%%%%%%%%%%%%%%%%%%%%%%%%%%%%%%
\subsection{Modified AdaTron learning}
%%%%%%%%%%%%%%%%%%%%%%%%%%%%%%%%%%%%%%%%%%%%%%%%%%%%%%%%%%%%%%%%%%%%
%%
%%
In the previous subsection, we found that the 
on-line AdaTron learning fails to obtain the zero 
residual error for finite $a$. 
In this subsection, we modify 
the AdaTron learning 
as $f={\Theta}(-T_{a}(v)S_{a}(u))h(u)l$ with 
\begin{eqnarray}
h(u)=\left\{
\begin{array}{rl}
a-u & \mbox{($u>\frac{a}{2}$)} \\
-u & \mbox{($-\frac{a}{2}<u<\frac{a}{2}$)} \\
-a-u & \mbox{($u<-\frac{a}{2}$)} 
\end{array}\right. 
\label{adat}
\end{eqnarray}
and see if the generalization ability 
of our non-monotonic system is improved. 
The motivation for the above 
choice comes from the optimization 
of the learning algorithm to be mentioned in the 
next section. 
Details of derivation of Eq. (\ref{adat}) 
are found in Appendix A.
Then the differential equation with 
respect to $R$ is obtained as follows. 
\begin{equation}
\frac{d R}{d \alpha}=
-\frac{R^{2}}{2}E_{\rm MA}(R)-RF_{\rm MA}(R)+G_{\rm MA}(R)
\label{drdamod}
\end{equation}
where $E_{\rm MA}(R)={\ll}h^{2}(u){\Theta}(-T_{a}(v)S_{a}(u)){\gg}$, 
$F_{\rm MA}(R)={\ll}uh(u){\Theta}(-T_{a}(v)S_{a}(u)){\gg}$ and 
$G_{\rm MA}(R)={\ll}vh(u){\Theta}(-T_{a}(v)S_{a}(u)){\gg}$. 
To see the asymptotic behavior of 
the generalization error, we evaluate 
the leading-order contribution as $R$ 
approaches $1$, $R=1-{\varepsilon}$, as 
\begin{eqnarray}
E_{\rm MA} & {\sim} & \frac{2\sqrt{2}}{\pi}(1+2{\Delta})
{\varepsilon}^{\frac{3}{2}}
\label{sime}\\
F_{\rm MA} & {\sim} & -\frac{2\sqrt{2}}{\pi}
\left(1+2(1-a^{2}){\Delta}\right)
{\varepsilon}^{\frac{3}{2}}
\label{simf} \\
G_{\rm MA} & {\sim} & \frac{4\sqrt{2}a^{2}{\Delta}}{\pi}
{\varepsilon}^{\frac{3}{2}}.
\label{simg} 
\end{eqnarray}
Substituting these expressions  into the differential 
equation (\ref{drdamod}), we obtain 
${\varepsilon}^{1/2}=\sqrt{2}\pi/(1+2{\Delta}){\alpha}^{-1}$ 
and the generalization error as 
\begin{equation}
{\epsilon}_{g}=\frac{\sqrt{2}(1+2{\Delta})}{\pi}{\varepsilon}^{\frac{1}{2}}
=\frac{2}{\alpha}.
\label{gemod}
\end{equation}
We should notice that the above 
result is independent of $a$ 
and the generalization ability of the student 
is improved by this  modification for all finite $a$. 
%%
%%
%%
%%
%%
%%%%%%%%%%%%%%%%%%%%%%%%%%%%%%%%%%%%%%%%%%%%%%%%%%%%%%%%%%%%%%
%%%%%%%%%%%%%%%%%%%%%%%%%%%%%%%%%%%%%%%%%%%%%%%%%%%%%%%%%%%%%%
\section{Optimized learning}
%%%%%%%%%%%%%%%%%%%%%%%%%%%%%%%%%%%%%%%%%%%%%%%%%%%%%%%%%%%%%%%
\subsection{Optimization of the learning rate }
%%%%%%%%%%%%%%%%%%%%%%%%%%%%%%%%%%%%%%%%%%%%%%%%%%%%%%%%%%%%%%%
%%%%%%%%%%%%%%%%%%%%%%%%%%%%%%%%%%%%%%%%%%%%%%%%%%%%%%%%%%%%%%%%
%%
%%
%%
%%
%%
In the present subsection, we improve the conventional perceptron learning 
by introducing a time-dependent learning late \cite{Ino1,Ino2}. 
We consider the next on-line dynamics; 
\begin{equation}
{\bf J}^{m+1}={\bf J}^{m}
-g(\alpha)\,{\Theta}(-T_{a}(v)S_{a}(u))S_{a}(u){\bf x}.
\label{optj}
\end{equation}

Using the same technique as in the previous section, 
we can derive the differential 
equations with respect to $l$ and $R$ as follows. 
\begin{eqnarray}
\frac{d l}{d \alpha} & = & 
\frac{1}{l}
\left[\frac{1}{2}g(\alpha)^{2}E(R)-g(\alpha)F(R)l
\right]
\label{optdlda}\\
\frac{d R}{d \alpha} & = & 
\frac{1}{l^{2}}
\left[-\frac{R}{2}E(R)g(\alpha)^{2}+g(\alpha)(F(R)R-G(R))l
\right] \nonumber \\
\mbox{} & {\equiv} & L(g(\alpha)).
\label{optdrda}
\end{eqnarray}
The optimal learning rate $g_{\rm opt}(\alpha)$ 
is determined so as to maximize $L(g(\alpha))$ 
to accelerate the increase of $R$.
We then find 
\begin{equation}
g_{\rm opt}=\frac{[F(R)R-G(R)]l}{RE(R)}.
\label{optrate}
\end{equation}
Substituting this expression into the above 
differential equations,  we obtain 
\begin{eqnarray}
\frac{d l}{d R} & = & 
-\frac{[F(R)R-G(R)][F(R)R+G(R)]l}{2R^{2}E(R)}
\label{optdlda2}\\
\frac{d R}{d l} & = & \frac{[F(R)R-G(R)]^{2}}
{2RE(R)}.
\label{optdrda2}
\end{eqnarray}
We can obtain the asymptotic form of ${\varepsilon}\,(=1-R)$, 
$l$ and ${\epsilon}_{g}$ with the same 
technique of analysis as in the previous section; 
\begin{eqnarray}
{\varepsilon} & = & 4
\left[
\frac{2\sqrt{2}(1+2{\Delta})}{(1-2{\Delta})^{2}}
\right]^{2}
{\alpha}^{-2}, \\
l & = & {\exp}
\left[
-16
\left(
\frac{1+2\Delta}{(1-2\Delta)^{2}}
\right)^{4}
{\alpha}^{-4}
\right],
\end{eqnarray}
and 
\begin{equation}
{\epsilon}_{g}=\frac{\sqrt{2}}{\pi}
(1+2\Delta)
\left[
\frac{2\sqrt{2}(1+2\Delta)}{(1-2{\Delta})^{2}}
\right]
{\alpha}^{-1}. 
\label{gemod2}
\end{equation}
Therefore, the generalization 
ability has been improved from ${\alpha}^{-1/3}$ 
for $g=1$ to ${\alpha}^{-1}$. 
The optimal learning rate  
$g_{\rm opt}(\alpha)$ behaves asymptotically as 
\begin{equation}
g_{\rm opt}=\frac{2\sqrt{2\pi}}{(1-2{\Delta})}
{\alpha}^{-1}
{\exp}
\left(-16
\left(
\frac{1+2\Delta}{(1-2\Delta)^{2}}
\right)^{4}
{\alpha}^{-4}
\right). 
\label{asmlr}
\end{equation}
The factor $F(R)R-G(R)$ of $g_{\rm opt}$ 
appearing in Eq.  (\ref{optrate}) 
is calculated by substituting 
$F(R)$ and $G(R)$ in Eqs. (\ref{frper}) and (\ref{grper}) 
as $F(R)R-G(R)=(1-R^{2})(1-2\Delta)/\sqrt{2\pi}$. 
Thus, at $a=a_{c}=\sqrt{2\log{2}}$, 
the optimal learning rate vanishes. 
Therefore our formulation 
does not work at $a=a_{c}$. 

As the optimal learning rate 
$g_{\rm opt}$ changes the sign at $a=a_{c}$,  
from the arguments in section III, 
we can see why the optimal learning rate can eliminate 
the anti-learning. 
%%%%%%%%%%%%%%%%%%%%%%%%%%%%%%%%%%%%%%%%%%%%%%%%%%%%%%%%%%%%%%%%%%%%%%%%%
%%%%%%%%%%%%%%%%%%%%%%%%%%%%%%%%%%%%%%%%%%%%%%%%%%%%%%%%%%%%%%%%%%%%%%%%%

In relation to this phenomenon at $a=\sqrt{2{\log}2}$, 
Van den  Broeck \cite{cris,cris2} recently investigated 
the same reversed-wedge perceptron which learns 
in the unsupervised mode from 
the distribution 
\begin{equation}
P(v)=2\frac{{\exp}(-\frac{v^{2}}{2})}
{\sqrt{2\pi}}
\left[
{\Theta}(v-a)+{\Theta}(v+a){\Theta}(-v)
\right]
\label{nondist}
\end{equation}
with $v=\sqrt{N}({\bf B}{\cdot}{\bf x})/|{\bf B}|$. 
For small ${\alpha}$, he found $R({\alpha})\,{\sim}\,\sqrt{\alpha<v>^{2}}$ 
for the optimal on-line learning, 
where $<{\cdots}>$ denotes the average over the distribution (\ref{nondist}).
Then he showed that at $a=\sqrt{2{\log}2}$, the distribution 
(\ref{nondist}) leads to $<v>=0$ and consequently 
$R({\alpha}){\equiv}0$.
From this result, he concluded that 
as long as $<v>=0$ holds, any kind of on-line learning necessarily fails and 
the corresponding learning curve 
has a plateau. 
It seems that a similar mechanism may lead to a 
failure of the optimal learning 
at $a=\sqrt{2{\log}2}$ in our model.
%%
%%
%%
%%%%%%%%%%%%%%%%%%%%%%%%%%%%%%%%%%%%%%%%%%%%%%%%%%%%%%%%%%%%%%%%%%%%%%%
%%%%%%%%%%%%%%%%%%%%%%%%%%%%%%%%%%%%%%%%%%%%%%%%%%%%%%%%%%%%%%%%%%%%%
%%
%%
%%
%%%%%%%%%%%%%%%%%%%%%%%%%%%%%%%%%%%%%%%%%%%%%%%%%%%%%%%%%%
%%%%%%%%%%%%%%%%%%%%%%%%%%%%%%%%%%%%%%%%%%%%%%%%%%%%%%%%%%%%
\subsection{Optimization of the weight function using the Bayes formula}
%%%%%%%%%%%%%%%%%%%%%%%%%%%%%%%%%%%%%%%%%%%%%%%%%%%%%%%%%%%%%%
%%%%%%%%%%%%%%%%%%%%%%%%%%%%%%%%%%%%%%%%%%%%%%%%%%%%%%%%%%%%%%
%%
%%
In this subsection we try another optimization 
procedure by Kinouchi and Caticha \cite{Kino}. 
We choose the optimal 
weight function $f(T_{a}(v),u)$ 
by differentiating the right hand side of Eq.  (\ref{drdagen}) 
with the aim to 
accelerate the increase of 
$R$
\begin{equation}
f^{*}=\frac{l}{R}(v-Ru).
\label{optf1}
\end{equation}
It is important to remember that 
$f^{*}$ contains some  unknown information 
for the student, namely, the 
local field of the teacher 
$v$. 
Therefore, we should 
average $f^{*}$ over a suitable 
distribution to erase $v$ from $f^{*}$. 
For this purpose, we transform 
the variables $u$ and $v$ to $u$ and $z$  
\begin{equation}
v=z\sqrt{1-R^{2}}+Ru.  
\label{trans}
\end{equation}
Then, the connected Gaussian distribution 
$P_{R}(u,v)$ is rewritten as 
\begin{eqnarray}
P_{R}(u,v)=\frac{1}{2\pi\sqrt{1-R^{2}}}
{\exp}(-\frac{u^{2}}{2})\,
{\exp}(-\frac{z^{2}}{2}).
\label{distri2}
\end{eqnarray}
We then obtain 
\begin{equation}
<f^{*}>
=\frac{\sqrt{1-R^{2}}}{R}l<z>
\label{fstar} 
\end{equation}
where $<\cdots>$ stands for the averaging over the 
variable $v$. 
Substituting this into the differential 
equation (\ref{drdagen}), we find 
\begin{equation}
\frac{d R}{d \alpha}=\frac{(1-R^{2})}{2R}
{\ll}<z>^{2}{\gg}.
\label{drdabayes2}
\end{equation}

Let us now calculate $<z>$. 
For this purpose, we use the 
distribution $P(z|y,u)$. 
This quantity means 
the posterior probability of $z$ 
when $y$ and $u$ are given, 
where we have set $y\,{\equiv}\,T_{a}(v)$. 
This conditional probability 
is rewritten by the Bayes formula 
\begin{equation}
P(z|y,u)=\frac{P(z)P(y|u,z)}{\int\,dz\,P(z)P(y|u,z)},
\label{bayes}
\end{equation}
from which we can calculate $<z>$ as 
\begin{eqnarray}
<z> & = & \int\,dz\,z\,P(z|y,u) \nonumber \\
\mbox{} & = & \frac{\int\,dz\,z\,P(z)\,P(y|u,z)}
{\int\,dz\,P(z)\,P(y|u,z)} \nonumber \\
\mbox{} & = & \frac{\int\,z\,Dz\,P(y|u,z)}{\int\,Dz\,P(y|u,z)}. 
\end{eqnarray}
Here $P(y|u,z)$ is given as  
\begin{eqnarray}
P(y|u,z) & = & y\,{\Theta}(z\sqrt{1-R^{2}}+Ru) \nonumber \\
\mbox{} & - & y\,{\Theta}(z\sqrt{1-R^{2}}+Ru-a) \nonumber \\
\mbox{} & + & y\,{\Theta}(-z\sqrt{1-R^{2}}-Ru-a) \nonumber \\
\mbox{} & + & \frac{1}{2}(1-y)
\label{condprob}
\end{eqnarray}
from the distribution $y=T_{a}(v)$.
Then, the denominator of Eq.  (\ref{bayes}) 
is calculated as 
\begin{eqnarray}
\int\,Dz\,P(y|u,z) & = & y\int\,Dz\,{\Theta}(z\sqrt{1-R^{2}}+Ru) \nonumber \\
\mbox{} & - & y\int\,Dz\,{\Theta}(z\sqrt{1-R^{2}}+Ru-a) \nonumber \\
\mbox{} & + & y\int\,Dz\,{\Theta}(-z\sqrt{1-R^{2}}-Ru-a)
+\frac{1}{2}(1-y) \nonumber \\
\mbox{} & {\equiv} & {\Omega}(y|u), 
\label{omega}
\end{eqnarray}
where ${\Omega}(y|u)$ means the posterior probability 
of $y$ when the local field of the student $u$ is given. 
As we treat the binary output teacher, we obtain 
from Eq. (\ref{omega})
\begin{equation}
{\Omega}({\pm}1|u)=H({\mp}\frac{Ru}{\sqrt{1-R^{2}}})
{\mp}H(\frac{a-Ru}{\sqrt{1-R^{2}}})
{\pm}H(\frac{a+Ru}{\sqrt{1-R^{2}}}). 
\end{equation}
In Figs. 6 ($R=0.5$) and 7,  
($R=0.9$), we plot ${\Omega}(+1|u)$ for the cases of 
$a=4.0, 2.0, 1.0$ and $a=0.5$. 
From these figures, we find that 
for any $a$  ${\Omega}(+1|u)$ 
seems to reach $(T_{a}(u)+1)/2$ as 
$R$ goes to $+1$. 
Using the same technique, we can calculate 
$\int\,Dz\,z\,P(y|u,z)$ and obtain 
\begin{equation}
\int\,Dz\,z\,P(y|u,z)=\frac{\sqrt{1-R^{2}}}{R}
\frac{\partial }{\partial u}
{\Omega}(y|u).
\label{averz}
\end{equation}
Substituting this into the right hand side of 
${d R}/{d \alpha}$, Eq. (\ref{drdabayes2}), we obtain 
\begin{equation}
\frac{d R}{d \alpha}=
{\ll}
-\frac{(1-R^{2})^{2}}{2R^{3}}
\left\{
\frac{\partial}{\partial u}
{\log}{\Omega}(y|u)
\right\}^{2}
+
\frac{(1-R^{2})^{3/2}z}{R^{2}}
\frac{\partial}{\partial u}{\log}
{\Omega}(y|u)
{\gg}, 
\end{equation}
where ${\ll}{\cdots}{\gg}$ stands for the averaging over 
the distribution $P(y,u)=\int\,{Dz}\,P(y|u,z)P(u)P(z)$. 
Performing this average, we finally obtain
\begin{equation}
\frac{d R}{d \alpha}=
\frac{(1-R^{2})}{4{\pi}R}
\int_{-\infty}^{\infty}\,Du\,{\Xi}_{a}(R,u)
\label{drdabayes}
\end{equation}
where 
\begin{eqnarray}
{\Xi}_{a}(R,a) & {\equiv} & 
\left[
{\exp}(-\frac{A_{1}^{2}}{2})
-{\exp}(-\frac{A_{2}^{2}}{2})
-{\exp}(-\frac{A_{3}^{2}}{2})
\right]^{2} \nonumber \\
\mbox{} & {\times} & 
\left[
\frac{1}{H(-A_{1})-H(A_{2})+H(A_{3})}
+\frac{1}{H(A_{1})+H(A_{2})-H(A_{3})}
\right]
\end{eqnarray}
and 
$A_{1}\,{\equiv}\,{Ru}/{\sqrt{1-R^{2}}}$, 
$A_{2}\,{\equiv}\,(a-Ru)/{\sqrt{1-R^{2}}}$, 
$A_{3}\,{\equiv}\,(a+Ru)/{\sqrt{1-R^{2}}}$. 
We plot the generalization error 
by numerically solving Eqs. (\ref{dldaper}),\,(\ref{drdaper}),\,
(\ref{optdlda2}),\,(\ref{optdrda2}), and (\ref{drdabayes}) 
for the cases of 
$a=\infty$ in Fig. 8  and $a=1.0$ in Fig. 9. 
From these figures, we see that 
for the both cases of $a=\infty$ and $a < \infty$, the generalization 
error calculated by the Bayes formula converges more 
quickly to zero than by 
the optimal learning rate $g_{\rm opt}(\alpha)$. 
%%
%%
%%
%%%%%%%%%%%%%%%%%%%%%%%%%%%%%%%%%%%%%%%%%%%%%%%%%%%%%%%%%%%%%%%%%%%%%%%%
%%%%%%%%%%%%%%%%%%%%%%%%%%%%%%%%%%%%%%%%%%%%%%%%%%%%%%%%%%%%%%%%%%%%%%%%%

Recently, Simmonetti and Caticha \cite{SC} 
introduced the on-line learning algorithm for 
the non-overlapping parity machine  with 
general number of nodes $K$.  
In their method, the weight vector of 
the student in each hidden unit is trained by the 
method in Ref. \cite{Kino}. 
In order to average over the internal fields of teacher 
in the differential equation with respect to the specific 
hidden unit $k$ of the student, 
they need the conditional probability which depends 
not only on the internal field of the unit $k$ but also on 
the internal field of the other units ($i{\neq}k$). 
This fact shows 
that their optimal algorithm is non-local.
In our problem, the input-output relation of the machine 
can be mapped to those of 
a single layer reversed-wedge perceptron. 
Therefore, it is not necessary for us to use the 
information about all units and our optimizing procedure 
leads to a local algorithm.  
%%
%%
%%
%%
%%%%%%%%%%%%%%%%%%%%%%%%%%%%%%%%%%%%%%%%%%%%%%%%%%%%%%%%%%%%%%%%%%%%%%%%%%%
%%%%%%%%%%%%%%%%%%%%%%%%%%%%%%%%%%%%%%%%%%%%%%%%%%%%%%%%%%%%%%%%%%%%%%%%%%

In order to investigate the 
performance of the Bayes optimization, 
we have calculated the asymptotic 
form of the generalization error from Eq. 
(\ref{drdabayes}) and the result is 
\begin{equation}
{\varepsilon}^{\frac{1}{2}}=\frac{2}{(1+2{\Delta})C{\alpha}}
\end{equation}
for ${\varepsilon}=1-R$,
where
\begin{equation}
C\,{\equiv}\,\frac{1}{{\pi}^{3/2}}\int_{-\infty}^{\infty}
dt\,\frac{{\exp}(-t^{2})}{H(t)}.
\end{equation}
The generalization error is then given by Eq. (\ref{asme1}) as 
\begin{equation}
{\epsilon}_{g}=\frac{2\sqrt{2\pi}}{\int_{-\infty}^{\infty}
dt\,{{\exp}(-t^{2})}
/{H(t)}}\,\,\,\frac{1}{\alpha}\,{\sim}\,0.883\,\,\,\frac{1}{\alpha}.
\end{equation}
This asymptotic form of the generalization error 
agrees with the result of Kinouchi and Caticha \cite{Kino}.
We notice that 
this form is independent of the width of the reversed wedge $a$.

We next mention the physical meaning 
of ${\Xi}_{a}(R,u)$ appearing in the differential 
equation (\ref{drdabayes}). 
As the rate of increase ${d R}/{d \alpha}$ is 
proportional to ${\Xi}_{a}(R,u)$, 
this quantity is regarded as 
the distribution of the gain  which determines the 
increase of $R$. 
Therefore, ${\Xi}_{a}(R,u)$ yields 
important information about 
the strategy to make queries. 
A query means to restrict the input signal to the 
student, $u$, to some subspace. 
Kinzel and Ruj$\acute{\rm a}$n suggested 
that if the student learns by the Hebbian learning algorithm 
from restricted inputs, 
namely, inputs lying on the subspace $u=0$, the prefactor of 
the generalization error becomes a half \cite{Kinzel}.
In the present formulation (\ref{drdabayes}), 
a query-making  can be incorporated by 
inserting appropriate delta functions in the 
integrand.
The learning process is clearly accelerated 
by choosing the peak position 
of ${\Xi}_{a}(R,u)$ as the location 
of these delta functions. 
In Fig. 10  we plot the distribution ${\Xi}_{a}(R,u)$ 
for $a=2.0$ (top) and $a=0.8$ (bottom). 
From these figures, we learn that for large $a$ ($=2.0$), 
the most effective example lies on the decision boundary 
($u=0$) at the  initial training stage (small $R$).
However, as the student learns, two different peaks appear
 symmetrically and in the final stage of training, 
the distribution has three peaks around $u=0$ and 
$u={\pm}a$. 
On the other hand, for small $a$ ($=0.8$), 
the most effective examples lie 
at the tails ($u={\pm}\infty$) for 
the initial stage. 
In the final stage, the distribution has two peaks 
around $u={\pm}a$. 
Therefore it is desirable to change the 
location of queries adaptively.
%%
%%
%%
%%%%%%%%%%%%%%%%%%%%%%%%%%%%%%%%%%%%%%%%%%%%%%%%%%%%%%%%%%%%%%%%%%%%%
%%%%%%%%%%%%%%%%%%%%%%%%%%%%%%%%%%%%%%%%%%%%%%%%%%%%%%%%%%%%%%%%%%%%%
\section{Conclusion}
%%%%%%%%%%%%%%%%%%%%%%%%%%%%%%%%%%%%%%%%%%%%%%%%%%%%%%%%%%%%%%%%%%%%%%
%%
%%
We have investigated the generalization abilities 
of a non-monotonic perceptron, 
which may also be regarded as a multilayer neural network, 
a parity machine, 
in the on-line mode. 
We first showed that the 
conventional perceptron and Hebbian learning 
algorithms lead to the perfect learning 
$R=1$ only when $a>a_{c}=\sqrt{2{\log}2}$. 
The same algorithms yield the 
opposite state $R=-1$ in the other 
case $a<a_{c}$. 
These algorithms have originally been designed 
having the simple perceptron ($a=\infty$) in mind, 
and thus are natural to give the opposite result for the 
reversed-output system ($a\,{\sim}\,0$).
In contrast, the 
conventional AdaTron learning algorithm 
failed to obtain the zero residual error for 
all finite values of $a$. 
For the unlearnable situation (where the structures 
of the teacher and student are different), 
Inoue and Nishimori reported 
that the AdaTron learning 
converges to the largest residual error among 
the three algorithms \cite{Ino2}.
It is interesting that 
the AdaTron learning algorithm is not useful 
even for the learnable situation. 

In order to overcome this difficulty, 
we introduced several modified versions of 
the conventional learning rules. 
We first introduced the time-dependent 
learning rate into the on-line 
perceptron learning and optimize it. 
As a result, the generalization error 
converges to zero in proportion to 
${\alpha}^{-1}$ except at 
$a=\sqrt{2{\log}2}$ where  
the learning rate becomes identically zero. 
We next improved the conventional AdaTron learning by 
modifying the weight function so that 
it changes according to the 
value of the internal potential $u$ of the student. 
By this modification, the generalization ability 
of the student dramatically improved 
and the generalization error 
converges to zero with an $a$-independent 
form, $2{\alpha}^{-1}$.

We also investigated a different type of 
optimization:
We first optimized 
the weight function $f(T_{a}(v),u)$ appearing in 
the on-line dynamics, not the rate $g$. 
Then, as the function $f$ contains 
the unknown variable $v$, 
we averaged it over the distribution of $v$ using 
the well-known technique of the Bayes statistics. 
This optimization procedure also 
provided other useful information 
for the student, namely, 
the distribution of most effective examples. 
Kinzel and Ruj$\acute{\rm a}$n \cite{Kinzel} reported that 
for the situation in which a simple perceptron 
learns from a simple perceptron (the $a=\infty$ case), the Hebbian 
learning with selected examples ($u=0$) leads to 
faster convergence of the generalization error than the 
conventional Hebbian learning. 
However, we have found that 
for finite values of $a$, the most effective 
examples lie  not only  on the boundary $u=0$ but also 
on $u={\pm}a$. 
Furthermore, we could learn that 
for small values of $a$ and at the initial stage of 
learning ($R$ small), 
the most effective examples lie on 
the tails ($u={\pm}\infty$). 
As the learning proceeds, the most effective examples 
change the locations to $u={\pm}a$. 
This information is useful for effective 
query constructions 
adaptively at each stage of learning.
%%
%%
%%
%%%%%%%%%%%%%%%%%%%%%%%%%%%%%%%%%%%%%%%%%%%%%%%%%%%%%%%%%%%%%
%%%%%%%%%%%%%%%%%%%%%%%%%%%%%%%%%%%%%%%%%%%%%%%%%%%%%%%%%%%%%
\acknowledgements 
%%%%%%%%%%%%%%%%%%%%%%%%%%%%%%%%%%%%%%%%%%%%%%%%%%%%%%%%%%%%%
%%%%%%%%%%%%%%%%%%%%%%%%%%%%%%%%%%%%%%%%%%%%%%%%%%%%%%%%%%%%%
%%
%%
We thank Professor Shun-ichi Amari for useful discussions. 
One of the authors (J.I.) was a partially supported by 
the Junior Research Associate Program of RIKEN. 
He also thanks Professor  C. Van den Broeck for useful 
discussion.
Y.K. was partially supported by a program ``Research for the 
future (RFTF)'' of Japan Society for the Promotion of Science.
The authors thank the referee for a very careful reading and 
a number of constructive comments.
%%%%%%%%%%%%%%%%%%%%%%%%%%%%%%%%%%%%%%%%%%%%%%%%%%%%%%%%%%%%%%%%%
%%%%%%%%%%%%%%%%%%%%%%%%%%%%%%%%%%%%%%%%%%%%%%%%%%%%%%%%%%%%%%%%%
\appendix
%%%%%%%%%%%%%%%%%%%%%%%%%%%%%%%%%%%%%%%%%%%%%%%%%%%%%%%%%%%%%%%%%
%%%%%%%%%%%%%%%%%%%%%%%%%%%%%%%%%%%%%%%%%%%%%%%%%%%%%%%%%%%%%
%%%%%%%%%%%%%%%%%%%%%%%%%%%%%%%%%%%%%%%%%%%%%%%%%%%%%%%%%%%
\section{Derivation of the weight function in 
the modified AdaTron 
learning algorithm}
%%%%%%%%%%%%%%%%%%%%%%%%%%%%%%%%%%%%%%%%%%%%%%%%%%%%%%%%%%%%%%
%%%%%%%%%%%%%%%%%%%%%%%%%%%%%%%%%%%%%%%%%%%%%%%%%%%%%%%%%
%%
In this appendix, we explain 
how we introduced the modified weight function 
${\Theta}(-T_{a}(v)S_{a}(u))h(u)l$
appearing in the AdaTron learning algorithm in Sec. IV B. 
From Eqs. (\ref{fstar}) and (\ref{averz}) in Sec. V, 
the weight function 
using the Bayes formula is written as 
\begin{equation}
<f^{*}>=\frac{1-R^{2}}{R^{2}}\,l\,\frac{\partial }{\partial u}
{\log}\,{\Omega}\,(y|u).
\end{equation}
As this expression contains the unknown parameter 
$R$ to  the student, we try to find   
the suitable learning weight function which agrees with the 
asymptotic form of $<f^{*}>$ in the limit of $R{\rightarrow}1$ 
\cite{Biehl94}. 
For this purpose, we investigate the asymptotic form of 
${\Omega}\,(y|u)$ as follows. 
We consider the cases of $T_{a}{\equiv}y=1$ 
and $y=-1$ separately.
\\
\\
(I)\,$y=1$\\
Using the relation $R=1-{\varepsilon}, 
{\varepsilon}{\rightarrow}0$, we find 
\begin{eqnarray}
{\Omega}(y|u) & = & H\left(
-\frac{Ru}{\sqrt{1-R^{2}}}
\right)
-H\left(
\frac{a-Ru}{\sqrt{1-R^{2}}}
\right)
+H\left(
\frac{a+Ru}{\sqrt{1-R^{2}}}
\right) \nonumber \\
\mbox{} & {\simeq} & \frac{1}{\sqrt{\pi}}
\left[
{\rm erfc}\left(\frac{-u}{2\sqrt{\varepsilon}}\right)
-{\rm erfc}\left(\frac{a-u}{2\sqrt{\varepsilon}}\right)
+{\rm erfc}\left(\frac{a+u}{2\sqrt{\varepsilon}}\right)
\right].
\label{asmomega1}
\end{eqnarray}
The asymptotic form of ${\Omega}\,(y|u)$ depends on the 
range of $u$. 
For $u>a$, the asymptotic form of 
${\Omega}\,(y|u)$ is 
\begin{eqnarray}
{\Omega}{\sim}\frac{1}{u-a}\sqrt{\frac{\varepsilon}{\pi}}
{\exp}\left(
-\frac{(u-a)^{2}}{4{\varepsilon}}
\right).
\end{eqnarray}
Therefore, $<f^{*}>/l=-(u-a)$.
Similarly, we find 
$<f^{*}>/l=0$ ($0<u<a$ and $u<-a$), 
$<f^{*}>/l=-u$ ($-a/2<u<0$) and 
$<f^{*}>/l=-(u+a)$ ($-a<u<-a/2$). 
\\
(II)\,$y=-1$ \\
Using the relation $R=1-{\varepsilon}$, we find for $u>a$ 
\begin{eqnarray}
{\Omega}{\sim}1-\frac{1}{u-a}\sqrt{\frac{\varepsilon}{\pi}}
{\exp}\left(
-\frac{(u-a)^{2}}{4{\varepsilon}}
\right).
\end{eqnarray}
Therefore, the weight function $<f^{*}>/l$ is $0$ 
asymptotically. 
Similarly, we find 
$<f^{*}>/l=0$ ($a/2<u<a$ and $-a<u<0$), 
$<f^{*}>/l=-u$ ($0<u<a/2$) and 
$<f^{*}>/l=-(a+u)$ ($u<-a$).

From the results of (I) and (II), we find the modified AdaTron 
learning algorithm as 
\begin{equation}
{\bf J}^{m+1}={\bf J}^{m}+{\Theta}(-T_{a}(v)S_{a}(u))
h(u)\,l\,{\bf x}
\end{equation}
where 
\begin{eqnarray}
h(u)=\left\{
\begin{array}{rl}
a-u & \mbox{($u>\frac{a}{2}$)} \\
-u & \mbox{($-\frac{a}{2}<u<\frac{a}{2}$)} \\
-a-u & \mbox{($u<-\frac{a}{2}$)} 
\end{array}\right. 
\end{eqnarray}
%%
%%
%%
%%
%%
%%%%%%%%%%%%%%%%%%%%%%%%%%%%%%%%%%%%%%%%%%%%%%%%%%%%%%%%%%%%%%%%%%
%%%%%%%%%%%%%%%%%%%%%%%%%%%%%%%%%%%%%%%%%%%%%%%%%%%%%%%%%%%%%%%%%%%

%%
%%
%%%%%%%%%%%%%%%%%%%%%%%%%%%%%%%%%%%%%%%%%%%%%%%%%%%%%%%%%%%%%%%%%%%%%%
%% Figure caption
%%%%%%%%%%%%%%%%%%%%%%%%%%%%%%%%%%%%%%%%%%%%%%%%%%%%%%%%%%
%%\Figures
%%
%%
\begin{figure}
\caption{
Generalization error as a function of the overlap 
$R$  for several values of $a$.
The student should be trained so that the overlap goes to $1$.
}
\end{figure}
\begin{figure}
\caption{
Trajectories of the $R$-$l$ flow for $a=\infty$.
All $R$-$l$ flows converge to the state of $R=1$ after 
infinite number of examples are represented. 
}
\end{figure}
\begin{figure}
\caption{
Trajectories of the $R$-$l$ flow for $a=0$.
All $R$-$l$ flows converge  to the state 
$R=-1$. Therefore, the corresponding generalization 
error does not converges to the ideal value of zero for this case. 
}
\end{figure}
%%
%%\begin{figure}
%%\caption{
%%Learning curves corresponding to Figs. 2 and 3.
%%}
%%\end{figure}
%%
%%
\begin{figure}
\caption{
Trajectories for the conventional AdaTron learning. 
Except for the case of $a=\infty$ and $a=0$ (overlapping), the trajectories converge to 
the state $l=0$.
}
\end{figure}
\begin{figure}
\caption{
Learning curves corresponding to Fig. 4. 
For the two cases of $a=\infty$ and $a=0$ (overlapping), 
the generalization errors converge to zero as ${\alpha}^{-1}$. 
However, for  the other cases, generalization errors converge to 
the finite value exponentially.
}
\end{figure}
\begin{figure}
\caption{
Shapes of ${\Omega}(+1|u)$ for $R=0.5$. 
}
\end{figure}
\begin{figure}
\caption{
Shapes of ${\Omega}(+1|u)$ for $R=0.8$.
We see that 
for any $a$ ${\Omega}(+1|u)$ 
seems to reach $(T_{a}(u)+1)/2$ as
$R$ goes to $+1$. 
}
\end{figure}
\begin{figure}
\caption{
Learning curves of perceptron, optimized perceptron and Baysian 
optimization algorithms for $a=\infty$. 
The Baysian optimization algorithm is the best among the three.
}
\end{figure}
\begin{figure}
\caption{
Learning curves of perceptron, optimized perceptron and Baysian 
optimization algorithms  for  $a=1.0$.
The Baysian optimization algorithm gives the best result 
among the three.
}
\end{figure}
\begin{figure}
\caption{
Distributions of the gain ${\Xi}_{a}(R,u)$ for $a=2.0$ (top) and 
$a=0.8$ (bottom).
The peak positions give the best place to make queries. 
}
\end{figure}
%%
%%
%%%%%%%%%%%%%%%%%%%%%%%%%%%%%%%%%%%%%%%%%%%%%%%%%%%%%%%%%%%%%%%%%%%%%%%
\end{document}